\documentclass[%
reprint,
 amsmath,amssymb,
 aps,
 prb,
 groupedaddress,
 superscriptaddress
]{revtex4-1}

\usepackage{graphicx}
\usepackage{dcolumn}
\usepackage{bm}
\usepackage{color}
\usepackage{float, subfigure}

\newcommand{\Tr}{\text{Tr}}
\newcommand{\dd}{\text{d}}
\newcommand{\neel}{N$\acute{\rm{e}}$el }

\begin{document}
\preprint{APS/123-QED}

\title{Finite temperature density matrix embedding theory}

\author{Chong Sun}
\email{csun2@caltech.edu}
\author{Ushnish Ray} 
\author{Zhi-Hao Cui}
\affiliation{Division of Chemistry and Chemical Engineering, California Institute of Technology, Pasadena, CA, 91125, USA}
\author{Miles Stoudenmire}
\affiliation{Center for Computational Quantum Physics, Flatiron Institute, New York, NY, 10010, USA}
\author{Michel Ferrero}
\affiliation{CPHT, CNRS, Ecole Polytechnique, Institut Polytechnique de Paris, Route de Saclay, 91128 Palaiseau, France}
\affiliation{Coll\`ege de France, 11 place Marcelin Berthelot, 75005 Paris, France}
\author{Garnet Kin-Lic Chan}
\email{garnetc@caltech.edu}
\affiliation{Division of Chemistry and Chemical Engineering, California Institute of Technology, Pasadena, CA, 91125, USA}

\date{\today}

\begin{abstract}

We describe a formulation of the density matrix embedding theory
at finite temperature. We present a generalization of the
ground-state bath orbital construction that embeds a mean-field finite-temperature density matrix up to a given order in the Hamiltonian, or the Hamiltonian
up to a given order in the density matrix. We assess the performance
of the finite-temperature density matrix embedding on the 1D Hubbard model both at
half-filling and away from it, and the 2D Hubbard model at half-filling,
comparing to exact data where available, as well as results from finite-temperature
density matrix renormalization group, 
 dynamical mean-field theory,
 and dynamical cluster approximations. 
The accuracy of finite-temperature
density matrix embedding appears comparable to that of the ground-state theory, with at most a modest increase in bath size,
and competitive with that of cluster dynamical mean-field theory.

\end{abstract}

\maketitle

\section{\label{sec:intro}Introduction}
The numerical simulation of strongly correlated electrons is key
to understanding the quantum phases that derive from
electron interactions, ranging from the Mott transition\cite{MottRMP1968,BullaPRL1999,BelitzRMP1994,QazilbashScience2007} to high 
temperature superconductivity\cite{AndersonScience1987,LakeScience2001,LakeNature2002}. Consequently, many numerical methods have been developed for this task.
In the setting of quantum lattice models, quantum
embedding methods\cite{SunACR2016}, such as dynamical mean-field theory (DMFT)\cite{KotliarRMP2006,GeorgesRMP1996,LichtensteinPRL2001,LichtensteinPRB2000,ZgidJCP2011} and density matrix embedding theory (DMET)\cite{Knizia2012,KniziaJCTC2013,WoutersJCTC2016,ZhengPRB2016,ZhengScience2017,BulikPRB2014,BulikJCP2014},
have proven useful in obtaining insights into complicated quantum phase diagrams.
These methods are based on an approximate mapping from the full interacting quantum lattice to a simpler
self-consistent quantum impurity problem, consisting of a few sites of the original lattice
coupled to an explicit or implicit bath. In this way, they avoid
treating an interacting quantum many-body problem in the thermodynamic limit.

The current work is concerned with the extension of DMET to finite temperatures.
DMET so far has mainly been applied in its ground-state formulation (GS-DMET), where it has achieved some success, particularly in applications to quantum phases where the order is associated with large unit cells~\cite{ZhengPRB2016,ZhengScience2017,ChenPRB2014}.
The ability to treat large unit cells at relatively low cost compared to other quantum embedding methods is due to the
computational formulation of DMET, which is based on modeling 
the ground-state impurity density matrix, a time-independent quantity
accessible to a wide variety of efficient quantum many-body methods.
Our formulation of finite-temperature DMET (FT-DMET) is based on the simple structure of GS-DMET,
but includes the possibility to generalize the bath so as to better capture the finite-temperature impurity density matrix.
Bath generalizations have previously been used to extend GS-DMET to the calculation
of spectral functions and other dynamical quantities~\cite{BoothPRB2015,fertitta2019energy}. Analogously to GS-DMET, since one only needs to
compute time-independent observables,
finite-temperature DMET can be paired with the wide variety of quantum impurity solvers which can provide the finite-temperature
density matrix.

We describe the theory of FT-DMET in  Section \ref{sec:theory}. In Section \ref{sec:results} we
carry out numerical calculations on the 1D and 2D Hubbard models, using exact diagonalization (ED) and the finite-temperature density matrix
renormalization group (FT-DMRG)\cite{FeiguinPRB2005} as quantum impurity solvers. We benchmark our results against those from the Bethe Ansatz in 1D, and
DMFT and the dynamical cluster approximation (DCA) in 2D, and also explore the quantum impurity derived \neel transition in the 2D Hubbard model.
We finish with brief conclusions about prospects for the method in \ref{sec:conclusions}.

\section{\label{sec:theory}Theory}

\subsection{Ground state DMET}\label{sec:gsdmet}

In this work, we exclusively discuss DMET in lattice models (rather than for ab initio simulations~\cite{WoutersJCTC2016,KniziaJCTC2013,BulikJCP2014,cui2019efficient}).
As an example of a lattice Hamiltonian, and one that we will use in numerical simulations,
we define the Hubbard model~\cite{HubbardPRS1963,GutzwillerPRL1963}, 
\begin{equation}\label{eq:theory-hubham}
\hat{H} = -t\sum_{\langle i,j\rangle,\sigma} \hat{a}^{\dagger}_{i\sigma}
\hat{a}_{j\sigma} - \mu \sum_{i,\sigma} \hat{a}^{\dagger}_{i\sigma}
\hat{a}_{i\sigma} + U\sum_{i}\hat{n}_{i\uparrow}\hat{n}_{i\downarrow}
\end{equation}
where $\hat{a}^{\dagger}_{i\sigma}$ creates an electron with spin $\sigma$
on site $i$ and $\hat{a}_{i\sigma}$ annihilates it;
$\hat{n}_{i\sigma} = \hat{a}^{\dagger}_{i\sigma}\hat{a}_{i\sigma}$;
$t$ is the nearest-neighbour (denoted $\langle i,j\rangle$) hopping amplitute, here set to $1$;
$\mu$ is a chemical potential; 
and $U$ is the on-site repulsion.

The general idea behind a quantum embedding method such as DMET
is to approximately solve the interacting problem in the large lattice by dividing the lattice into small
fragments or impurities~\cite{SunACR2016}. (Here we will assume the impurities are non-overlapping).
The main question is how to treat the coupling and entanglement between the impurities.
In DMET, other fragments around a given impurity are modeled by a set of bath orbitals.
The bath orbitals are constructed to exactly reproduce the entanglement between
the impurity and environment when the full lattice is treated at a mean-field level (the so-called ``low-level'' theory).
The impurity together with its bath orbitals then constitutes a small embedded quantum impurity problem, 
which can be solved with a ``high-level'' many-body method. The low-level lattice wavefunction 
 and the high-level embedded impurity wavefunction are made approximately
consistent, by enforcing self-consistency of the single-particle density matrices
of the impurities and of the lattice. This constraint is implemented by introducing a static correlation potential on
the impurity sites into the low-level theory.


To set the stage for the finite-temperature theory, in the following we briefly
recapitulate some details of the above steps in the GS-DMET formulation. In particular, we discuss
how to extract the bath orbitals, how to construct the embedding hamiltonian, and how to carry out the
self-consistency between the low-level and high-level methods. 
Additional details for the GS-DMET algorithm can be found in several articles~\cite{Knizia2012,ZhengPRB2016,WoutersJCTC2016}, including the review in Ref.~\onlinecite{WoutersJCTC2016}.

\subsubsection{DMET bath construction}

We first assume that we have a mean-field (``low-level'') wavefunction $|\Phi\rangle$
defined on the full lattice of $L$ sites; $|\Phi\rangle$ is an eigenstate of a quadratic lattice Hamiltonian $\hat{h}$.
Then, given an impurity $x$ defined over $L_x$ sites, DMET relies on the lemma that $|\Phi\rangle$ can be rewritten in the form
\begin{align}
  \label{eq:theory-mfwf}
  |\Phi\rangle = |\Phi_\text{emb}\rangle |\Phi_\text{core}\rangle
  \end{align}
where $|\Phi_\text{emb}\rangle$ lives in a Hilbert space of the $L_x$ sites of the impurity and an additional $L_x$ bath orbitals
defined in its environment; $|\Phi_\text{core}\rangle$ is a product state defined in
the complement of the impurity and environment Hilbert spaces.
This basic result states that the entanglement between the impurity and its environment contained in $|\Phi\rangle$ is entirely
captured within the small impurity-bath Hilbert space.

The single-particle density matrix $D^\Phi$ obtained from $|\Phi\rangle$
contains all information on the correlations in $|\Phi\rangle$.
Thus the bath orbitals can be defined from this density matrix.
We consider the impurity-environment block $D^{\Phi}_{\text{imp-env}}$ ($D_{ij}$ for $i \in x, j \notin x$) of dimension $L^{x} \times (L-L^x)$.
Then taking the thin SVD
\begin{equation}
D^{\Phi}_{\text{imp-env}} = U\lambda B^{\dagger}
\end{equation}
the columns of $B$ specify the bath orbitals in the lattice basis.
The bath space is thus a function of the density matrix, denoted $B(D)$.

\subsubsection{Embedding Hamiltonian}

After obtaining the bath orbitals, we construct the embedded Hamiltonian of the quantum impurity problem. 
In GS-DMET, there are two ways to do so, the
interacting bath formulation and the non-interacting bath formulation. The conceptually simplest approach
is the interacting bath formulation. In this case, we project
the interacting lattice Hamiltonian $\hat{H}$ into the space of
the impurity plus bath orbitals, defined by the projector $\hat{P}$, i.e. the embedded Hamiltonian
is  $\hat{H}_\text{emb} = \hat{P}\hat{H}\hat{P}$. 
$\hat{H}_\text{emb}$  in general contains non-local interactions involving the bath orbitals, as they are non-local
orbitals in the environment. From the embedded Hamiltonian, we compute the high-level
ground-state impurity wavefunction,
\begin{align}
  \hat{H}_\text{emb} |\Psi\rangle = E |\Psi\rangle
  \end{align}
If $\hat{H}$ were itself the quadratic lattice Hamiltonian $\hat{h}$, then
then $\Psi = \Phi$ and 
\begin{equation}\label{eq:theory-samegs}
\hat{P}\hat{h}\hat{P} |\Phi\rangle = E |\Phi\rangle
\end{equation}
Another way to write Eq.~(\ref{eq:theory-samegs}) for a mean-field state is
\begin{align}\label{eq:theory-samegs2}
  [{P}{h}{P}, {P}{D}^\Phi{P}] = 0
  \end{align}
where $h$ denotes the single-particle Hamiltonian matrix and $P$ is the single-particle
projector into the impurity and bath orbitals. 
These conditions imply that the  lattice Hamiltonian and the embedded Hamiltonian $\hat{H}_\text{emb}$ share
the same ground-state at the mean-field level, which is the basic approximation in GS-DMET.

In the alternative non-interacting bath formulation, interactions on the bath are
approximated by a quadratic correlation potential (discussed below). This formulation retains the same exact embedding property as the interacting
bath formulation for a quadratic Hamiltonian.
In practice,  both formulations give similar results in the Hubbard model\cite{BulikPRB2014,WuJCP2019}, and the choice between the two
depends on the available impurity solvers; the interacting bath formulation generates non-local two-particle interactions in the bath
that not all numerical implementations can handle.
In this work, we use the interacting bath formulation in the 1D Hubbard model where an ED solver is used.
In the 2D Hubbard model, we use the non-interacting bath formulation, where both ED and FT-DMRG solvers are used.
This latter choice is because the cost of treating non-local interactions in FT-DMRG is relatively high (and we
make the same choice with ED solvers to keep the results strictly comparable).

\subsubsection{Self-consistency}

To maintain self-consistency between the ground-state of the lattice mean-field $|\Phi\rangle$,  
and that of the interacting embedded Hamiltonian $|\Psi\rangle$, we introduce
a quadratic correlation potential $\hat{u}$ into $h$, i.e.
\begin{align}
  \hat{h} \to \hat{h} + \hat{u}
  \end{align}
where $\hat{u}$ is constrained to act on sites in the impurities, i.e. $\hat{u} = \sum_x \hat{u}^x$. To study magnetic order, we choose
the form
\begin{align}
  \hat{u}^x = \sum_{ij \in x, \sigma \in \{ \uparrow,\downarrow\}} u^x_{i j\sigma} a^\dag_{i\sigma} a_{j\sigma} 
\end{align}
The coefficients $u^x_{ij\sigma}$ are adjusted to match the density
matrices on the impurity that are evaluated from the low-level wavefunction $|\Phi\rangle$
and from the high-level embedded wavefunction $|\Psi\rangle$. In this work, we
only match the single-particle density matrix elements of the impurity (impurity only matching~\cite{WoutersJCTC2016}).
Note also that we will only be considering translationally invariant systems, and thus $\hat{u}^x$ is the same
for all impurities.

\subsection{Ground-state expectation values}

\label{sec:gsexpect} 
Ground-state expectation values are evaluated from the density matrices of each high-level impurity wavefunctions
$|\Psi^x\rangle$. Since there are multiple impurities (in a translationally invariant system, these
are constrained to be identical) an expectation value is typically assembled from
the multiple impurity wavefunctions using a democratic partitioning~\cite{WoutersJCTC2016}. For example, given two sites $i$, $j$, where $i$
is part of impurity $x$ and $j$ is part of impurity $y$, 
\begin{align}
  \langle a^\dag_i a_j \rangle = \frac{1}{2}[\langle \Psi^x | a^\dag_i a_j | \Psi^x\rangle +\langle \Psi^y | a^\dag_i a_j | \Psi^y\rangle]
\end{align}
Note that the {\it pure bath} components of the high-level wavefunctions, e.g. $\langle \Psi^x | a^\dag_i a_j |\Psi^x\rangle$ for $i, j \notin x$
do not contribute to the DMET expectation values.
The democratic partitioning means an individual impurity embedding contributes the correct amount to a global expectation values so long as the impurity wavefunction produces
correct expectation values for operators that act on the impurity alone, or the impurity and bath together.

\subsection{Finite temperature DMET}\label{sec:ftdmet}

Our formulation of FT-DMET follows the same rubric as the ground-state theory: a low-level
(mean-field-like) finite-temperature
density matrix is defined for the lattice; this is used to
obtain a set of bath orbitals to define the impurity problem;
a high-level finite-temperature
density matrix is calculated for the embedded impurity; and
self-consistency is carried out between the two via a correlation potential.
The primary difference lies in the properties of the bath, which we focus on below,
as well as in the appearance of quantities such as the entropy, which are
formally defined from many-particle expectation values.

\label{subsec:ftdmet}
\subsubsection{Finite temperature bath construction}

In GS-DMET, the bath orbital construction is designed to be exact
if all interactions are treated at the mean-field level, giving rise
to the commuting condition for the projected single-particle density matrix
and projected Hamiltonian in Eq.~(\ref{eq:theory-samegs2}).
In general, we can look for a finite-temperature bath construction that preserves a similar property.
As pointed out in Sec.~\ref{sec:gsexpect}, the DMET embedding is still exact
for single-particle expectation values if the embedded projected single-particle
density matrix produces the correct expectation values in the impurity and impurity-bath
sectors, due to the use of the democratic partitioning. We aim to satisfy this slightly relaxed condition.


The finite temperature single-particle density matrix of a quadratic Hamiltonian $\hat{h}$
is given by the Fermi-Dirac function 
\begin{equation}\label{eq:theory-fdfull}
{D}(\beta) = \frac{1}{1 + e^{({h}-\mu)\beta}}
\end{equation}
where $\beta = 1/k_B T$ ($k_B$ is the Boltzmann constant, $T$ is the temperature). In the following, we fix $k_B = 1$, thus
$\beta = 1/T$. If we could find an embedding directly analogous to the ground-state construction, we would obtain a projector ${P}$, such that
the embedded density matrix ${P} {D}{P}$ is the Fermi-Dirac function of the embedded quadratic Hamiltonian, i.e.
${P} h {P}$, i.e.
\begin{equation}\label{eq:theory-fdemb}
{P} {D} {P} = \frac{1}{1 + e^{({P}{h}{P}-\mu)\beta}}
\end{equation}
However, unlike in the ground-state theory, the non-linearity of the exponential function means that
Eq.~(\ref{eq:theory-fdemb}) can only be satisfied 
exactly if ${P}$ projects back into the full lattice basis. Thus a bath orbital construction at
finite temperature is necessarily always approximate, even for quadratic Hamiltonians.

Nonetheless, one can choose the bath orbitals to reduce the error between the l.h.s. and r.h.s. in
Eq.~(\ref{eq:theory-fdemb}). First, we require that the equality is satisfied only
for the impurity only, and impurity-environment, blocks of $D$, following the relaxed requirements
of the democratic partitioning.  Second, we require the equality to
be satisfied only up to a finite order $n$ in $h$, i.e.
\begin{align}
\label{eq:fdemb_nth}
[P D P ]_{ij} = \left[\frac{1}{1 + e^{({P}{h}{P}-\mu)\beta}}\right]_{ij} + O(h^n) \quad i \in x,j \notin x 
\end{align}
Then there is a simple algebraic construction of the bath space  as (see Appendix for a proof)
\begin{align}
\label{eq:hbath}
\span\{B(h) \oplus B(h^2) \oplus B(h^3) \ldots B(h^n) \}
\end{align}
where $B(h^k)$ is the bath space derived from $h^k$, $k = 1, ..., n$. Note that each order of $h$ adds $L_x$ bath orbitals to the total
impurity plus bath space.

We can alternatively choose the bath to preserve the inverse relationship between the density matrix and Hamiltonian,
\begin{align}
[P h P ]_{ij} = \mathrm{inverseFD}(P D P) + O(D^n) \quad \mathrm{not}\ i,j \notin x 
\end{align}
where $\mathrm{inverseFD}$ is the inverse Fermi-Dirac function, and the bath space is
then given as
\begin{align}
\label{eq:dbath}
\span\{B(D) \oplus B(D^2) \oplus B(D^3) \ldots B(D^n) \}
\end{align}
The attraction of this construction is that the lowest order corresponds to the standard GS-DMET bath construction.

The above generalized bath constructions allow for the introduction of an unlimited number of bath sites (so long as the total number
of sites in the embedded problem is less than the lattice size). Increasing the size of the embedded problem (hopefully) increases the
accuracy of the embedding, but it also increases the computational cost. However, an alternative way to increase
accuracy is simply to increase the number of impurity sites. Which strategy is better is problem dependent,
and we will assess both in our numerical experiments.

\subsubsection{Thermal observables}

The thermal expectation value of an observable $\hat{O}$ is defined as
\begin{equation}\label{eq:theory-ftexpval}
\langle \hat{O}(\beta)\rangle = \Tr\left[\hat{\rho}(\beta)\hat{O}\right]
\end{equation}
Once $\hat{\rho}(\beta)$ is obtained from
 the high-level impurity calculation, for observables based on low-rank (e.g. one- and two-)particle
reduced density matrices, we evaluate Eq.~(\ref{eq:theory-ftexpval}) using the democratic partitioning formula for
expectation values in Sec.~\ref{sec:gsexpect}. 

We will also, however, be interested in the entropy per site, which is a many-particle expectation value.
Rather than computing this directly as an expectation value, 
we will obtain it by using the thermodynamic relation
$\frac{\dd S}{\dd E} = \beta$, and
\begin{equation} 
S(\beta_0) = S(0) + \int_{E(0)}^{E(\beta_0)} \beta(E) \dd E 
\end{equation}
 where $\beta_0$ is the desired inverse temperature, and $S(0) = \ln 4$.






\section{\label{sec:results}Results}

\subsection{Computational details}

We benchmarked the performance of FT-DMET in the 1D and 
2D Hubbard models as a function of $U$ and $\beta$. For the 1D Hubbard model, we compared our FT-DMET results
to exact solutions from the thermal Bethe Ansatz \cite{TakahashiPRB2002}.
For the 2D Hubbard model, the FT-DMET results were compared to DCA and 
 DMFT
 results~\cite{MF_DMFT2d,maier2005,kunes,jarrellPRB,jarrellEPL,LeBlancPRX2015}. We used large DMET mean-field lattices with periodic boundary conditions  (240 sites in 1D, $24 \times 24$ sites in 2D).
We used exact diagonalization (ED) and finite temperature DMRG (FT-DMRG)
as impurity solvers. 
There are two common ways to carry out finite temperature DMRG calculations: the purification method~\cite{FeiguinPRB2005} and the
minimally entangled typical thermal states (METTS) method~\cite{StoudenmireNJP2010}.
In this work, we used the purification method implemented with the ITensor
 package\cite{ITensor} as the FT-DMRG impurity solver, as well as to provide the finite lattice reference data in Fig.\ref{fig:1ddopping}.
In the 1D Hubbard model, we used ED exclusively and the interacting bath formulation of DMET, while in the 2D Hubbard model,
we used ED for the 4 impurity, 4 bath
calculations, and FT-DMRG for the 4 impurity, 8 bath calculations, both within the non-interacting bath formulation.
FT-DMRG was carried out using 4th order Runge-Kutta time evolution. 
To denote different calculations with different numbers of impurity and bath orbitals, we use the notation $InBm$, where $n$
denote the number of impurity sites and $m$ the number of bath orbitals.

\subsection{1D Hubbard model}\label{sec:1dhub}

The 1D Hubbard model is an ideal test system for FT-DMET as its thermal properties
can be exactly computed via the thermal Bethe ansatz. We thus use it to assess various choices
within the FT-DMET formalism outlined above.

\begin{figure}
\includegraphics[width=0.46\textwidth]{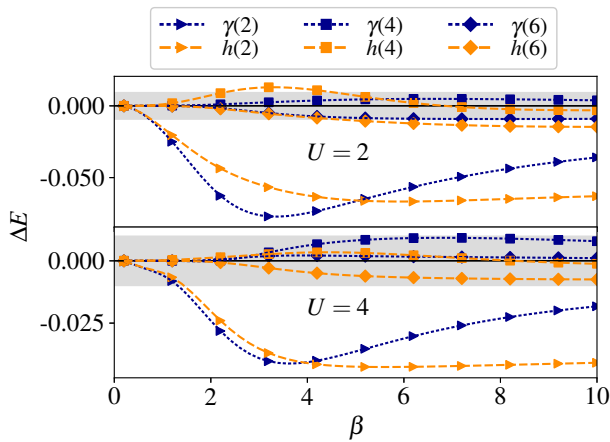}
\caption{Color online. Error in energy per site (units of $t$)  of FT-DMET 
     for the 1D Hubbard model at $U=2$ and $U=4$ (2 impurity sites and 
     half-filling) with bath orbitals generated via the 
       density matrix $\gamma$ (Eq.~(\ref{eq:dbath})) (blue lines) 
     or lattice Hamiltonian $h$ (Eq.~(\ref{eq:hbath})) (orange lines)
     as a function of inverse temperature $\beta$.  The number in 
    the parentheses denotes the number of bath orbitals. 
    The grey area denotes the ground state error with 2 impurity orbitals.
    }\label{fig:hbath-vs-dbath}
\end{figure}

We first compare the relative performance of the two proposed bath constructions, generated via the Hamiltonian in Eq.~(\ref{eq:hbath}) or
via the density matrix in Eq.~(\ref{eq:dbath}). In Fig.~\ref{fig:hbath-vs-dbath} we show the
error in the energy per site (measured from the thermal Bethe ansatz) for $U=2, 4$ and half-filling for these two choices. (The behaviour
for other $U$ is similar).
Using 4 bath sites, the absolute error in the energy is comparable to that of the ground-state calculation (which uses 2 bath sites)
over the entire temperature range.
Although the Hamiltonian
construction was motivated by the high temperature expansion of the density matrix, the density matrix construction appears to perform well
 at both low and high temperatures.
Consequently, we use the density matrix derived bath in the subsequent calculations.


\begin{figure}
    \centering
    \includegraphics[width=0.48\textwidth]{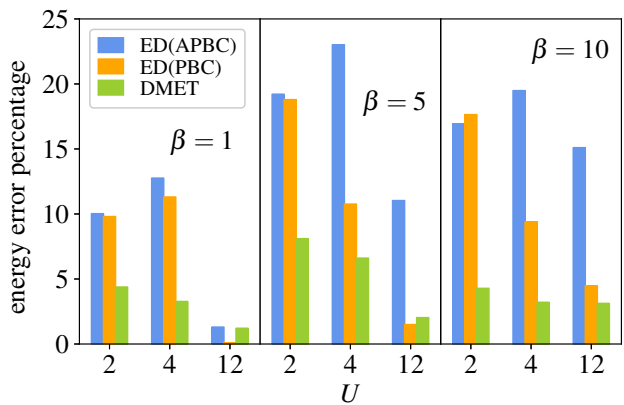}
    \caption{Color online. Percentage error of the FT-DMET (with 2 impurity sites and 2 bath orbitals) energy per site vs. ED (4 sites) on a non-embedded cluster with PBC and APBC boundary conditions
    for the 1D Hubbard model at various $U$ and $\beta$.} \label{fig:dmet-vs-ed}
\end{figure}

We next examine the effectiveness of the density matrix bath construction in removing the 
the finite size error of the impurity. As a first test, 
in Fig.~\ref{fig:dmet-vs-ed} we compare the energy error obtained with FT-DMET and  $I2B2$ with
a pure ED calculation with 4 impurity sites ($I4$) and periodic (PBC) or antiperiodic (APBC) boundary
conditions, at various $U$ and $\beta$. For weak ($U=2$) to moderate ($U=4$) coupling, FT-DMET shows a significant improvement
over a finite system calculation with the same number of sites, reducing the error by a factor of $\sim 2-6$ depending on the $\beta$, thus
demonstrating the effectiveness of the bath. The maximum FT-DMET energy error is 8.1, 6.6, 3.1\% for $U=2, 4, 8$.
At very strong couplings, the error of the finite system ED with PBC approaches that of FT-DMET. This is because both the finite size error
and the effectiveness of the DMET bath decrease as one approaches the atomic limit.

\begin{figure}
    \centering
    \includegraphics[width=0.42\textwidth]{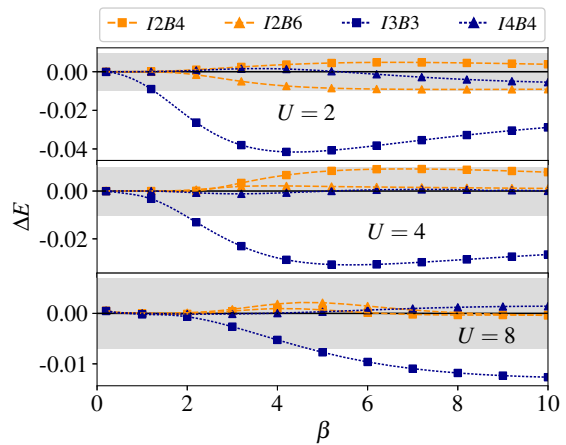}
     \caption{Color online. Percentage error of the FT-DMET
      energy per site of the 1D Hubbard model at half-filling
     as a function of impurity and bath size. $InBm$ denotes $n$ impurity sites and $m$ bath orbitals.
Increasing impurity (blue lines); increasing bath (orange lines). The grey band depicts the ground state error with $2$ impuriy sites and $2$ bath orbitals.}
     \label{fig:impvsbath}
\end{figure}

\begin{figure}
     \centering
     \includegraphics[width=3.5in]{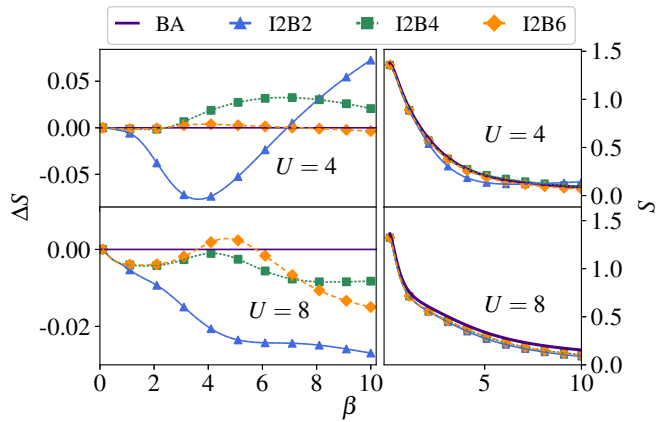}
     \caption{Color online. Error of the FT-DMET entropy per 
              site of the 1D Hubbard model at half-filling
              as a function of the number of bath sites. 
              The right patches shows the real values of the entropy.
               }

     \label{fig:1dES}
\end{figure}

As a second test, in Fig.~\ref{fig:impvsbath} we compare increasing the number of impurity sites versus increasing
the number of bath orbitals generated in Eq.~(\ref{eq:dbath}) for various $U$ and $\beta$. Although complex behaviour
is seen as a function of $\beta$, we roughly see two patterns. For certain impurity sizes, (e.g. $I4$) it can be slightly
more accurate to use a larger impurity with an equal number of bath sites, than a smaller impurity with a larger number of bath sites.
(For example, at $U=8$, one can find a range of $\beta$ where $I4B4$ gives a smaller error than $I2B6$). However, there
are also some impurity sizes which perform very badly; for example $I3B3$ gives a very large error, because the (short-range) antiferromagnetic
correlations do not properly tile between adjacent impurities when the impurities are of odd size. Thus, due to these
size effects convergence with impurity size is highly non-monotonic, but  increasing the bath size (by including more terms
in Eq.~(\ref{eq:dbath})) is less prone to strong odd-size effects.
The ability to improve the quantum impurity by simply increasing the number of bath sites,
is expected to be particularly relevant in higher-dimensional lattices such as the 2D Hubbard model, where ordinarily to
obtain a sequence of clusters with related shapes
it is necessary to increase the impurity size by large steps. Nonetheless, convergence with bath size is also not strictly monotonic,
as also illustrated in Fig.~\ref{fig:1dES}, where we see the error in the $I2B4$ entropy can sometimes be less
than that of $I2B6$ for certain ranges of $\beta$. For the largest embedded problem $I2B6$, the maximum
error in the entropy is $4\times 10^{-3}$, $2\times 10^{-2}$ for $U =4, 8$.


\begin{figure}
     \centering
     {\includegraphics[width=3.2in]{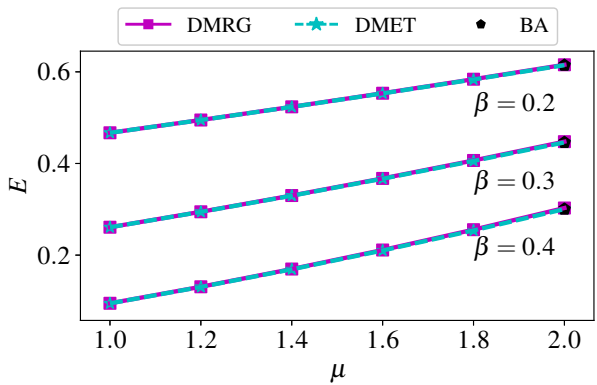}}
     \caption{Color online. Energy per site (units of $t$) of a 16-site 
     Hubbard chain with periodic boundary condition at $U=4$ as a function
     of the chemical potential $\mu$ at various $\beta$ values. 
     The difference between the DMRG and DMET energies per site is $0.01-0.8\%$.
     Solid lines: DMRG energies; dashed lines: DMET energies; pentagon: Bethe Ansatz.}
     \label{fig:1ddopping}
\end{figure}

The preceding calculations were all carried out at half-filling. Thus, 
in Fig.~\ref{fig:1ddopping} we show FT-DMET calculations on the 1D Hubbard model away from half-filling at $U=4$.
We chose to simulate a finite Hubbard chain of 16-sites with PBC in order to readily generate  numerically exact reference data
using FT-DMRG (using a maximum bond dimension of $2000$, and an
imaginary time step of $\tau = 0.025$). The agreement between the FT-DMRG energy per site and
that obtained from the thermal Bethe ansatz can be seen at half-filling, corresponding to a chemical potential $\mu=2$.
We see excellent agreement between FT-DMET and FT-DMRG results across the full range
of chemical potentials, and different $\beta$, suggesting that FT-DMET works equally well for doped system as
well as for undoped systems.

\subsection{2D Hubbard model}\label{sec:2dhub}

\begin{figure}
     \centering
     \includegraphics[width=3.3in]{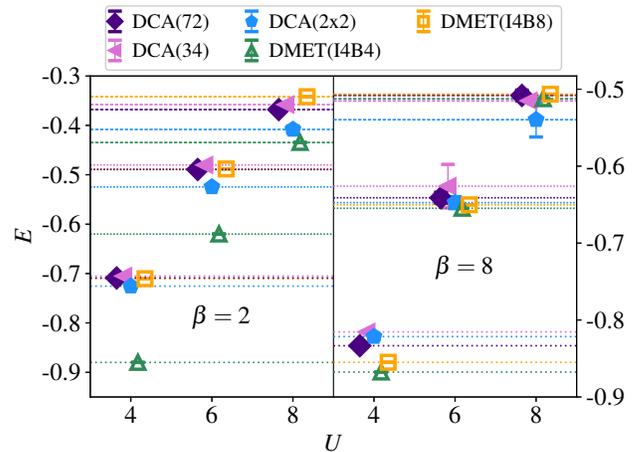}
     \caption{Color online. Energy per site versus U (units of $t$) of the 
     2D Hubbard model at half-filling with FT-DMET($2\times2$ cluster with 
     4 and 8 bath orbitals), 
     DCA(34, 72 and $2\times 2$ site clusters).
     }\label{fig:2dE}
     \end{figure}

The 2D Hubbard model is an essential model of correlation physics in materials.
We first discuss the accuracy of FT-DMET for the energy 
of the 2D Hubbard model at half-filling, shown in Fig.~\ref{fig:2dE}. The FT-DMET calculations
are performed on a $2\times 2$ impurity, with 4 bath orbitals ($I4B4$) (green diamond markers) 
and 8 bath orbitals ($I4B8$)  (red triangular markers). The results are compared
to DCA calculations with clusters of size $34$ (orange circle markers), $72$
(blue square markers)~\cite{LeBlancPRX2015}, and $2\times 2$  (light blue hexagon markers) (computed for this work). 
The DCA results with the size $72$ cluster can be considered here to represent the thermodynamic limit.
The 
DCA($2\times 2$) 
data provides an opportunity to assess the relative contribution of the FT-DMET embedding to the finite size error; in particular one can compare the 
difference between FT-DMET and DCA(72) to the difference between 
DCA($2\times2$) and DCA(72).
Overall, we see that the FT-DMET energies with 
8 bath orbitals are in good agreement with the DCA(72) energies across the different $U$ values, and that the accuracy is slightly
better on average than that of
DCA($2\times2$). 
The maximum error in the $I4B8$ impurity compared to thermodynamic limit extrapolations of the
DCA energy~\cite{LeBlancPRX2015} is found at $U=4$ and is in the range of 1-2\%,
comparable to errors observed in ground-state DMET at this cluster size (e.g. the error in GS-DMET at $U=4$ and $U=8$ is 0.3\% and 1.8\% respectively).
In the $\beta = 8$ case, the FT-DMET calculations with two different bath sizes give very similar results; at low temperature, the 
bath space constructed by the FT procedure is similar to that of the ground state, and the higher order bath sites do not contribute
very relevant degrees of freedom. Thus even the smaller bath achieves good accuracy in the low temperature FT-DMET calculations. 

\begin{figure}[h]
\subfigure[]{\includegraphics[width=3.1in]{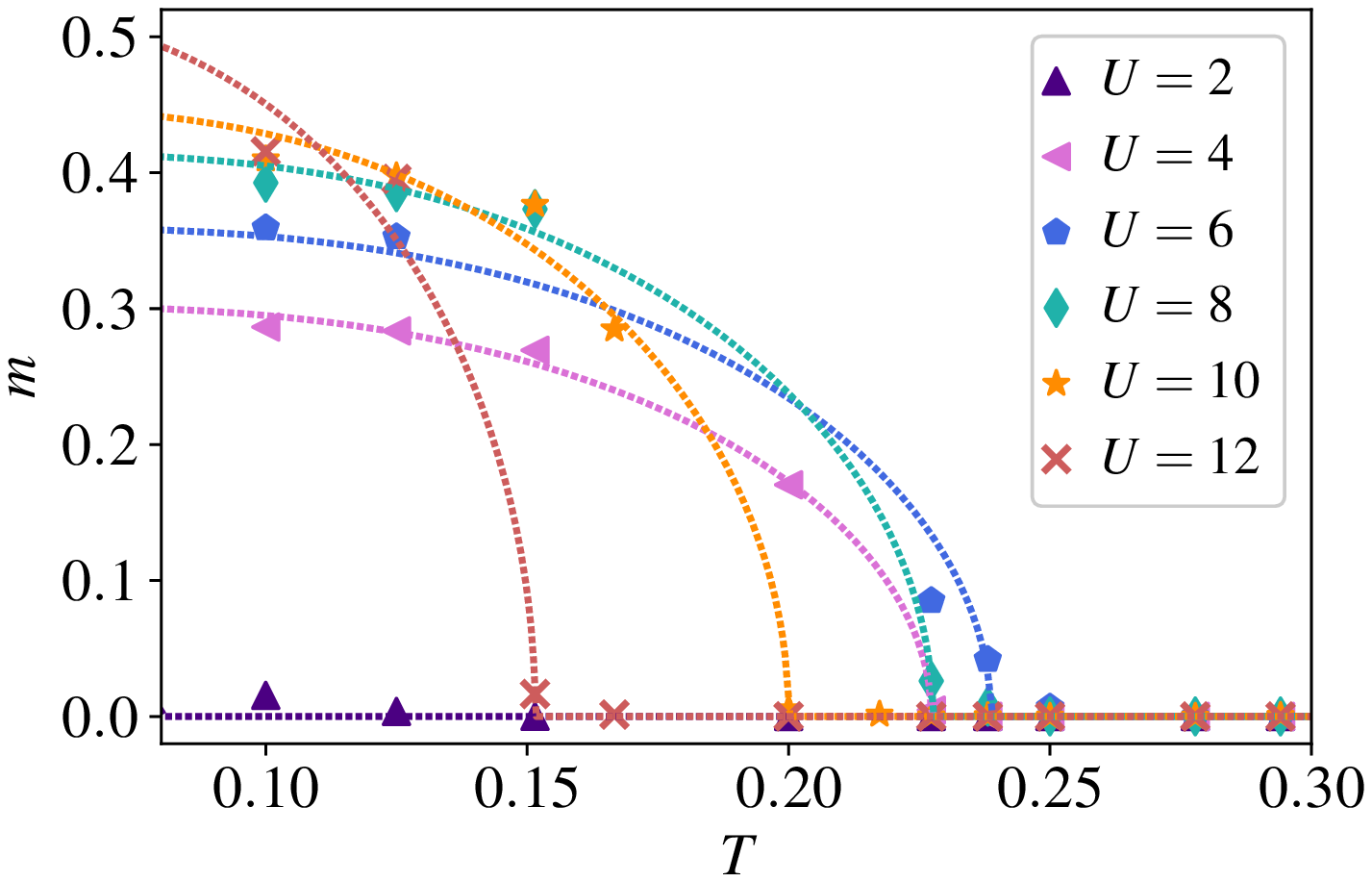}}
\subfigure[]{\includegraphics[width=3.1in]{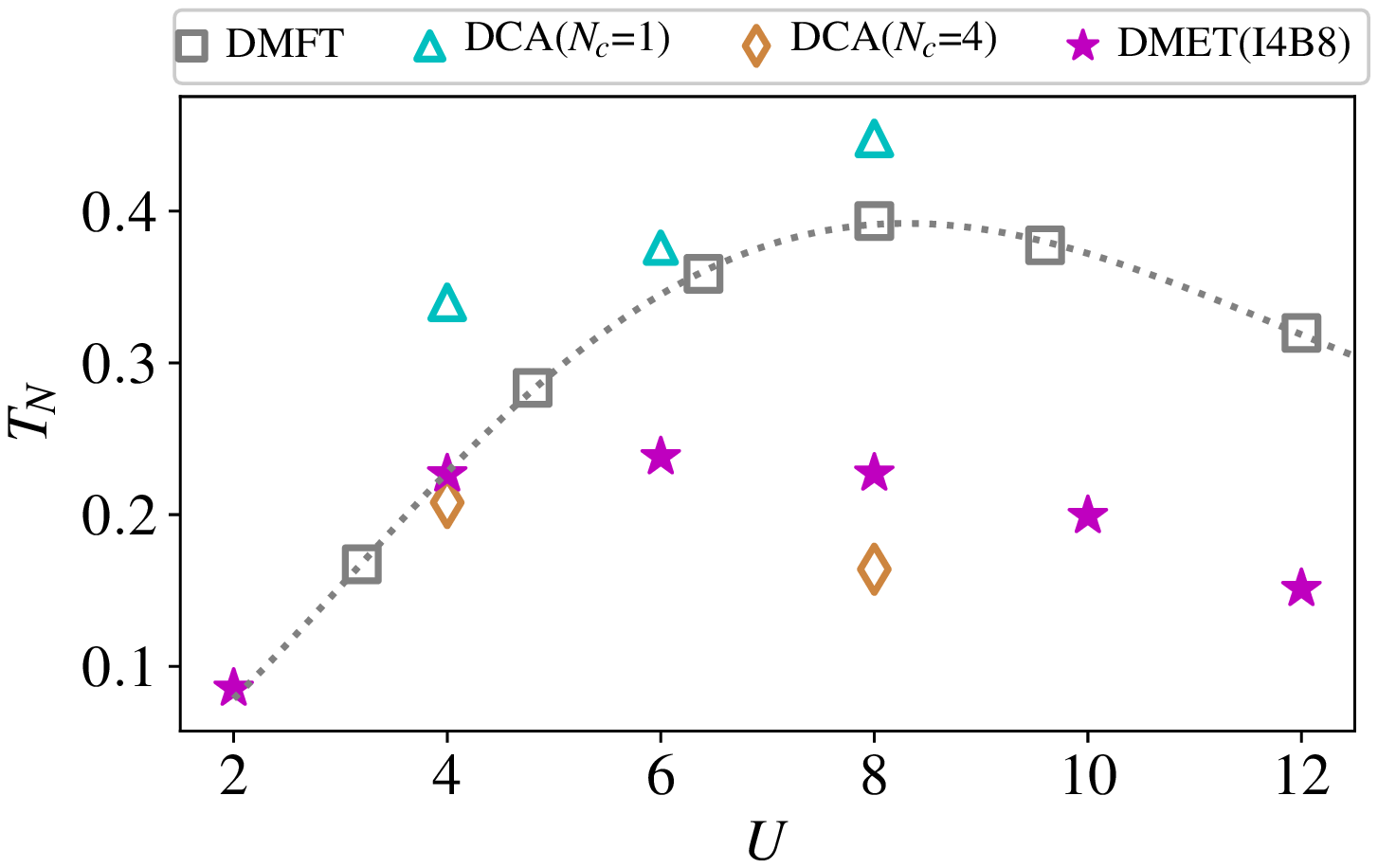}}
\caption{Color online. \neel  transition for the 2D Hubbard model within quantum impurity simulations.
 (a) Antiferromagnetic moment $m$ as a function of $T$ with various $U$ values (units of $t$);
(b) \neel  temperature $T_N$ calculated with FT-DMET, single-site DMFT and DCA.  DMFT data is taken from Ref.~\onlinecite{kunes}, DCA/NCA data for $U=4$ 
 is taken from Ref.~\onlinecite{maier2005}, DCA/QMC data for $U=6$ is 
 taken from Ref.~\onlinecite{jarrellPRB}, and DCA/QMC data for $U=8$ is 
 taken from Ref.~\onlinecite{jarrellEPL}.}\label{fig:2dlinemag}

\end{figure}

A central phenomenon in magnetism is the finite-temperature \neel  transition.
In the thermodynamic limit, the 2D Hubbard model does not exhibit a true \neel transition,
but shows a crossover~\cite{mermin}. However, in finite quantum impurity calculations,
the crossover appears as a phase transition at a nonzero \neel temperature. 
Fig.~\ref{fig:2dlinemag}(a)
shows the  antiferromagnetic moment $m$ calculated as $m = \frac{1}{2L_x}\sum_{i}^{L_x}|n_{i\uparrow}-n_{i\downarrow}|$ as a function of temperature $T$ for various 
$U$ values. As a guide to the eye, we fit the data to a mean-field 
magnetization function $m = a\tanh\left(bm/T\right)$,
where $a$ and $b$ are parameters that depend on $U$. The FT-DMET calculations are performed with a $2\times 2$ impurity and $8$ bath orbitals,
using a finite temperature DMRG solver  with maximal bond dimension $M = 600$ and time step $\tau = 0.1$. With this, the error in $m$ from the solver
is estimated  to be less than 10$^{-3}$.
$m$ drops sharply to zero as $T$ is increased
signaling a \neel   transition. The \neel  temperature $T_N$ is taken as the point of intersection
of the mean-field fitted line with the $x$ axis; assuming this form of the curve, the uncertainty in $T_N$ is invisible
on the scale of the plot.
The plot of $T_N$ versus $U$ is shown in Fig.~\ref{fig:2dlinemag}(b), showing that
 the maximal $T_N$ occurs at $U=6$.
Similar $T_N$ calculations on the 2D Hubbard model with single site DMFT~\cite{kunes} and DCA\cite{jarrellPRB,jarrellEPL, maier2005} are
also shown in Fig.~\ref{fig:2dlinemag}(b) for reference. Note that the difference in the DMFT results~\cite{kunes} and single-site DCA (formally equivalent to DMFT)~\cite{jarrellPRB,jarrellEPL}
likely arise from the different solvers used.
The behaviour of $T_N$ in our $2\times 2$ FT-DMET calculations
is quite similar to that of the 4-site DCA cluster.
In particular, we see in DCA that the
$T_N$ values obtained from calculations with a single-site cluster ($N_c=1$)
are higher than the $T_N$ values obtained from calculations with a 
4-site cluster ($N_c = 4$). 

\begin{figure}
\includegraphics[width=3in]{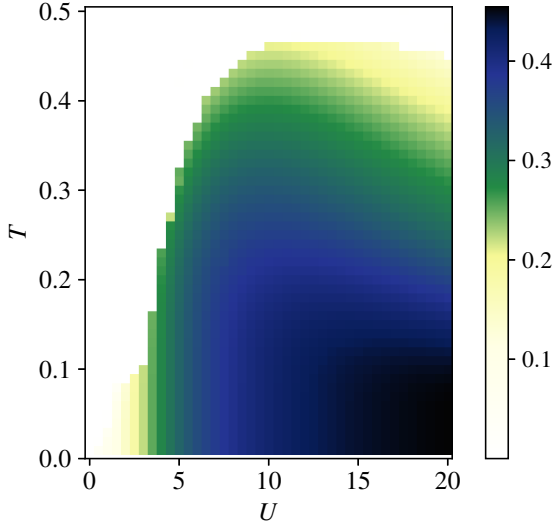}
\caption{Color online. 2D Hubbard antiferromagnetic moment (color scale) as a function of $T$ and $U$ (units of $t$) in FT-DMET ($2\times 2$ impurity, 4 bath orbitals.)}\label{fig:2dpd4b}
\end{figure}

An alternative visualization of the \neel transition in FT-DMET is shown in 
Fig.~\ref{fig:2dpd4b}. The FT-DMET calculations here were performed with 
a $2\times 2$ impurity and $4$ bath orbitals using an ED solver.
Though less quantitatively accurate than the $8$ bath orbital simulations, these FT-DMET calculations still capture
the qualitative behavior of the \neel  transition. Focusing on the dark blue
region of the phase diagram, one can estimate the maximal $T_N$ to occur near $U\approx 9$, an increase
over the maximal \neel temperature using the $8$ bath orbital impurity model. This increase
in the maximal $T_N$ appears similar to that which happens when moving from a 4-site cluster to 1-site cluster in DCA
in Fig.~\ref{fig:2dlinemag}.

\section{\label{sec:conclusions}Conclusions}
To summarize, we have introduced a finite temperature formulation
of the density matrix embedding theory (FT-DMET). This temperature
formulation inherits most of the basic structure of the ground-state
density matrix embedding theory, but modifies the bath construction
so as to approximately reproduce the mean-field finite-temperature density matrix.
From numerical assessments on the 1D and 2D Hubbard model, we conclude
that the accuracy of FT-DMET is comparable to that of its ground-state counterpart, with
at most a modest increase in size of the embedded problem. From the limited comparisons, it also
appears to be competitive in accuracy with cluster dynamical mean-field theory for the same sized cluster.
Similarly to ground-state DMET, we expect FT-DMET to be broadly applicable
to a wide range of model and ab initio problems of correlated electrons at finite temperature~\cite{ZhengScience2017,cui2019efficient}.

\begin{acknowledgements}
  This work was supported by the US Department of Energy
  via Award No. SC0018140. Additional support for GKC was provided
  by the Simons Foundation via the Simons Collaboration on the Many-Electron Problem, and via the Simons Investigator program.
 DCA(2x2) calculations were performed using HPC resources from GENCI (Grant No. A0070510609).
\end{acknowledgements}

\appendix
\section{Proof of the finite temperature bath formula}
\newcommand{\V}{V^{\dagger}}
\newcommand{\PP}{P^{\dagger}}
\newcommand*{\QEDB}{\hfill\ensuremath{\square}}%

Let $M$ be an arbitrary $N\times N$ full rank square matrix, and $Q_k$ be the $Q$ derived from the QR decomposition of the first $n$ columns of $M^k$, i.e., $M^k[:,:n] = Q_k R_k$, with $k = 0, 1, ..., K$. Let $S$ ($|S| < N$) be a space spanned by $\{Q_0, Q_1, ..., Q_K\}$, and $P$ be the projector onto $S$. The following equality holds
\begin{equation}\label{eq:2prove}
    P^{\dagger}M^lP[:,:n] = (P^{\dagger}MP)^l[:,:n], \hspace{0.2cm} l \leq K+1
\end{equation}

We prove the statement by mathematical induction. First write $M$ in the following form
\begin{equation}
    M = \begin{bmatrix}
    A & B \\
    C & D 
    \end{bmatrix}
\end{equation}
where $A$ and $B$ are the first $n$ rows of $M$, $A$ and $C$ are the first $n$
 columns of $M$. 
The projector has the form
\begin{equation}
    P = \begin{bmatrix}
        I & 0\\
        0 & V
        \end{bmatrix}
\end{equation}
where $I$ is an $n\times n$ matrix, and $V$ is an $(N-n)\times (K-1)n$ matrix with $(K-1)n < (N-n)$. The columns of $V$ are derived from the QR decomposition of $M^k[n:, :n]$, $k = 1, ..., K$ and then orthogonalized.
We can write $V$ in the form
\begin{equation}
    V = \begin{bmatrix} V_1 & V_2 & \cdots & V_K \end{bmatrix}
\end{equation}
where $V_k$ is from the QR decomposition of $M^k[n:, :n]$. $\PP M P$ has the 
form
\begin{equation}
\PP M P = \begin{bmatrix}
        A & BV\\
        \V C & \V DV
        \end{bmatrix}
\end{equation}

The mathematical induction consists of two parts:

(i) We start with $l=2$. The first $n$ columns of $P^{\dagger}M^2P$ and $(P^{\dagger}MP)^2$ are
\begin{equation}
\begin{split}
   P^{\dagger}M^2P[:,:n] &=  \begin{bmatrix}
    A^2 + BC \\ \V CA + \V DC
    \end{bmatrix}\\
    (P^{\dagger}MP)^2[:,:n] &= \begin{bmatrix}
    A^2 + BV\V C \\ \V CA + \V DV\V C
    \end{bmatrix}
\end{split}
\end{equation}
The two are equal when 
\begin{equation}\label{eq:VVC}
V\V C = V\V (VR) = V I R = VR = C
\end{equation}
which is true since $V$ is the $Q_1$ from the QR decomposition of $C$. (Note that $\V V = I$, but $V\V \neq I$).
 Therefore, Eq.~(\ref{eq:2prove}) holds for $l=2$ when $K \geq 1$. 

(ii) Now let us inspect Eq.~(\ref{eq:2prove}) for the $l$th order, assuming that Eq.~(\ref{eq:2prove}) holds for the $(l-1)$th order, i.e. $\PP M^{l-1}P = (\PP M P)^{l-1}$. Let 
\begin{equation}
    M^{l-1} = \begin{bmatrix}
    W & X \\
    Y & Z \\
    \end{bmatrix}
\end{equation}
and $M^l = MM^{l-1}$ has the form
\begin{equation}
    M^l = \begin{bmatrix}
    AW+BY  &  AX+BZ\\
    CW+DY  &  CX+DZ
    \end{bmatrix}
\end{equation}
and 
\begin{equation}
    \PP M^{l-1}P = (\PP MP)^{l-1} = \begin{bmatrix}
    W  &  XV\\
    \V Y  &  \V Z V
    \end{bmatrix}
\end{equation}
One can prove that $CW$ and $C$ share the same $Q$ space from the QR decomposition: let $C = QR$, then $CW = QRW$, where $R$ and $W$ are square matrices; we then perform another QR decomposition of $RW$, $RW = U\tilde{R}$, where $U$ is a unitary matrix, then $CW = \tilde{Q}\tilde{R}$ with $\tilde{Q} = QU$. Therefore, $Q$ and $\tilde{Q}$ span the same space. 

The first $n$ columns of $\PP M^lP$ and $(\PP M P)^l$ are
\begin{equation}
    \PP M^l P[:,:n]  = \begin{bmatrix}
    AW + BY \\
    \V CW + \V DY 
    \end{bmatrix}
\end{equation}

\begin{equation}
\begin{split}
    (\PP M P)^l[:,:n] =& \left((\PP M P)(\PP M P)^{l-1}\right)[:,:n]  \\
=& \begin{bmatrix}
     AW + BV\V Y\\
    \V CW + \V DV\V Y 
    \end{bmatrix}
\end{split}
\end{equation}
Since $V$ contains $V_{l-1}$, which is derived from the QR decomposition of $Y$, we have $V\V Y  = Y$ as in Eq.~(\ref{eq:VVC}). 

Combining (i) and (ii) we then see that Eq.~(\ref{eq:2prove}) holds for the $l$th order with $K\geq l-1$ for $\forall l$. \QEDB

\bibliography{reference}

\end{document}